\documentclass[conference,a4paper]{IEEEtran}
\usepackage[utf8]{inputenc}
\usepackage{amsmath}
\usepackage{amsfonts}
\usepackage{amssymb}
\usepackage[english]{babel}
\usepackage{graphicx}
\usepackage{comment}
\usepackage{cite}
\usepackage{cmap}
\usepackage{setspace}
\usepackage{xcolor}
\usepackage{multicol}
\usepackage{float}
\usepackage{tabularx,ragged2e,booktabs}
\newtheorem{definition}{Definition}
\newcolumntype{L}{>{\RaggedRight\arraybackslash}X} 
\renewcommand{\vec}[1]{\mathbf{#1}}

\begin{document}

\title{ Upper and Lower Bounds on Bit-Error Rate for Convolutional Codes
\thanks{}
}
\author{\IEEEauthorblockN{Anastasiia Kurmukova}\\
\IEEEauthorblockN{Institute for Information\\
Transmission Problems\\
Russian Academy of Sciences,\\
National Research University\\
Higher School of Economics, \\
Skoltech\\
Moscow, Russia \\
Anastasiia.Kurmukova@skoltech.ru}\\
\and
\IEEEauthorblockN{Fedor Ivanov}\\
\IEEEauthorblockN{National Research University\\
Higher School of Economics, \\
Institute for Information\\
Transmission Problems\\
Russian Academy of Sciences\\
Moscow, Russia \\
fivanov@hse.ru}\\
\and
\IEEEauthorblockN{Victor Zyablov}\\
\IEEEauthorblockN{Institute for Information\\
Transmission Problems\\
Russian Academy of Sciences\\
Moscow, Russia \\
zyablov@iitp.ru}}

\maketitle

\begin{abstract}

In this paper, we provide a new approach to the analytical estimation of the bit-error rate (BER) for convolutional codes for Viterbi decoding in the binary symmetric channel (BSC). The expressions we obtained for lower and upper BER bounds are based on the active distances of the code and their distance spectrum. The estimates are derived for convolutional codes with the rate $R=\frac{1}{2}$ but can be easily generalized for any convolutional code with rate $R=\frac 1n$ and  systematic encoder. The suggested approach is not computationally expensive for any crossover probability of BSC channel and convolutional code memory, and it allows to obtain precise estimates of BER. 

\end{abstract}


\section{Introduction}

Convolutional codes proposed by Elias in \cite{elias1955coding} have been studied for a long time. Although convolutional codes are inferior in error correction performance to such codes as LDPC \cite{8972062} or Polar \cite{ecp_yang} even for short data packets, these codes can significantly reduce bit-error rate (BER) for a given signal-noise ratio (SNR) \cite{7998249}. Both Polar and LDPC codes have no such property \cite{7998249}. Thus, convolutional codes can be called Error-reducing Codes (ERC) \cite{spielman1996linear}. Apart from convolutional codes only several classes of codes are ERC: for instance, Hamming codes \cite{rurik2016hamming}, some classes of LDPC codes with low-density generator matrix (LDGM) \cite{liu2010error}, repeated code constructions \cite{6814087}.

The property of BER reduction for a given SNR is very important for inner codes in any concatenated schemes. This is a major reason why convolutional codes as ERC codes are widely used in different concatenated and generalized concatenated schemes as woven codes \cite{host2002woven}, woven turbo codes \cite{freudenberger2000woven}, parallel concatenated codes \cite{benedetto1996design} e. t. c. Also the recursive systematic convolutional encoder has gained great popularity in turbo codes \cite{turbo_rsc}, \cite{turbo_rsc2}. It also shows great performance in concatenation with other linear code, for example, in parallel scheme with systematic polar code (SPC) with iterative decoding as in \cite{conv_polar_parallel}.

All of the above allows us to conclude that BER performance is the most important characteristic of convolutional codes. Of course, for a given convolutional code this characteristic can be evaluated by Monte-Carlo simulation. But this approach is inappropriate for the cases when very low BER is required. These cases correspond to ultra-high rate code constructions with very long constituent codes. Moreover, careful BER estimation allows to predict the performance of concatenated constructions and optimize parameters of outer codes for a given inner convolutional code.


Thus some analytical methods should be suggested to estimate BER performance of convolutional codes. There are several known approaches for this aim. In paper \cite{2819} simple Markov chains based models are used to evaluate BER of convolutional codes for short constraint lengths on very noisy channels. But simulation results provided in this paper show that for low input probability of error the proposed estimation are rather imprecise. Some modification of the technique proposed in \cite{2819} were suggested in \cite{chiaraluce1997technique}, where authors proposed to represent particular convolutional code as reduced Markov chain. In this case Viterbi decoding \cite{viterbi_bounds} can be considered as a special transition process between states of Markov chain. BER estimation for punctured convolutional codes was considered in \cite{6401021}. Finally, in paper \cite{bocharova2012closed} authors show how to estimate BER for any convolutional code both for quantized AWGN (additive white gaussian noise) and BSC (binary symmetric) channels. But the proposed technique is rather difficult in implementation and has high complexity for convolutional codes with large memory. 

In this paper, we use rather simple low complexity technique to estimate BER for convolutional codes, when codewords are transmitted over BSC and received words are decoded by Viterbi algorithm. Our approach is based on the distance spectrum of active distances of convolutional codes \cite{isita_burst}, \cite{my_redundancy}, that was previously used to estimate FER of convolutional codes \cite{isita_fer}. We derive both lower and upper bounds on BER and compare the analytical estimates we obtained with real simulations. 

\section{Distance Properties of Convolutional Codes}

\subsection{Convolutional codes with recursive encoder}

In this article, we consider convolutional codes of rate $R=\frac{1}{2}$ with recursive systematic encoder for simplicity. At the same time, the results we obtained can be generalised for convolutional codes of rate $R=\frac{1}{n}$, $n>1$, $n\in\mathbb{N}$. 

Binary symmetric channel is assumed as a communication channel. Viterbi decoder is assumed as a decoding algorithm.


The generator matrix for convolutional codes with systematic encoder with feedback is given with a division of polynomials:
\begin{eqnarray}
    \textbf{G}(D) = 
    \begin{pmatrix}
        1 & \frac{\textbf{g}^{(2)}}{\textbf{g}^{(1)}}  \label{genmatr}
    \end{pmatrix} \ ,
\end{eqnarray}
where generator polynomials $\textbf{g}^{(l)}(D) = g_{0}^{(l)} + g_{1}^{(l)} D + g_{2}^{(l)} D^2 +...+ g_{m}^{(l)} D^m$, $g_{i}^{(l)} \in \{0,1\} $, for $l = 1,2$. Here $m$ is a code memory, the code rate is $\frac{1}{2}$ and the code is systematic since the first polynomial equals to $1$. The feedback structure of the encoder corresponds to the denominator in the generator matrix. The generator polynomials have to be co-prime and same degree. The encoder state at each time moment can be written with $m$ register bits. The state also can be given as a decimal number $s$: $0\leq s < 2^m$.

The codeword of convolutional code can be written with a polynomial of an indeterminate $D$: $\vec{v}(D) = \vec{v}_0 + \vec{v}_1 D + \vec{v}_2 D^2 + ... $, $\vec{v}_i \in \{0, 1\}^2$, $i \in \mathbb{N} \cup \{0\}$, where a sequence of output $2$-tuples is $(\vec{v}_0, \vec{v}_1, \vec{v}_2, ...)$. The codeword is semi-infinite and can be also uniquely defined by a sequence of input information bits and the initial encoder state.

It is common practice to use trellis for a convolutional code description. Code trellis can provide a visual representation for the codeword as a sequence of the encoder states and transitions between them. Nodes of the trellis represent states of convolutional code encoder. Edge between two states exists if and only if there is a corresponding transition between these states. The transition from state $s_t$ at time moment $t$ to  state $s_{t+1}$ at time moment $t+1$ corresponds to some input bit and some output tuple from the encoder. Each codeword has unique corresponding sequence of output tuples, sequence of trellis states and sequence of input bits. Thus, each codeword can be written as a path in trellis (a sequence of the encoder states). Trellis representation is important for understanding the distance properties of convolutional codes.

\begin{figure}[htbp]
\centerline{\includegraphics[scale=0.3]{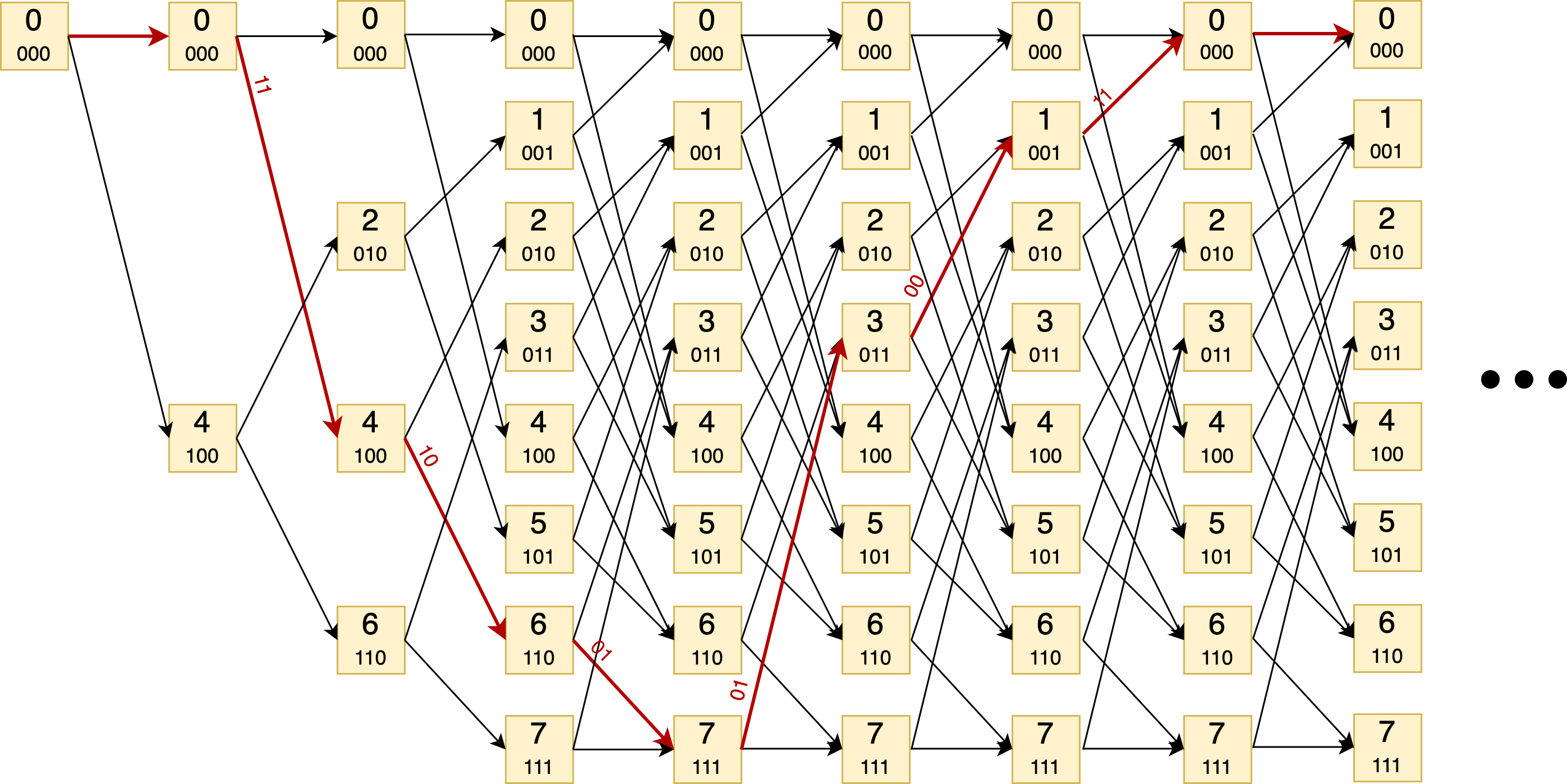}}
\caption{Trellis for the systematic code $(13,17)$}
\label{trellis1317}
\end{figure}

Here we provide an example in Fig. \ref{trellis1317} for the recursive systematic convolutional encoder of memory $m = 3$ with generator polynomials $\textbf{g}^{(1)} = 1 + D + D^3$, $\textbf{g}^{(2)} = 1 + D +  D^2 + D^3$ that can be written in octal form: $(13, 17)$. In Fig. \ref{trellis1317} there is a path highlighted in red corresponding to the beginning of one of the codewords. For a detailed explanation of the trellis concept, see \cite{viterbi_bounds}.

\subsection{Active Distances and Distance Spectrum of Active Distances}

Except a free (minimal) distance, convolutional codes have other important distance characteristics that determine the probability of erroneous decoding. Due to the linearity of convolutional code, free distance can be defined as the smallest weight of a non-zero codeword. 

It is known from the theory of linear codes that the probability of error is influenced by the number of words with the minimal weight. For block codes code spectrum is usually considered to estimate frame error probability. 

We suggest to analyse the error correcting performance of convolutional codes using active distances that were defined in \cite{act_dist}. For calculating a weight spectrum of convolutional code, a column distance can be used.  It does not increase with length after corresponded path in trellis has merged with allzero one. In active distances only those paths that do not pass two consecutive zero states are taken into account. Let us define active distances more accurately.


Further, we use the following notations: a codeword $\vec{v}(D)$ of the convolutional code denoted with corresponding sequence of the output tuples $(\vec{v}_0, \vec{v}_1, \vec{v}_2, \dots)$, $\vec{v}_i \in \{0, 1\}^2$, $i \in \mathbb{N} \cup \{0\}$, or a sequence of input bits $(u_0, u_1, u_2, \dots)$, or a sequence of trellis states $\textbf{s} = (s_{1}, s_{2}, \dots)$, where $s_{i} \in \{0,1,...,2^m-1\}$, $i \in \mathbb{N}$. The initial trellis state is usually considered as a zero: $s_0 = 0$.

First, we define a special subset of codewords $\mathcal{C}_f$ of convolutional code $\mathcal{C}$. The codewords from this subset have finite non-zero path on trellis that begins right after initial zero state and does not pass the two consecutive consecutive zero trellis states before merging with allzero path.

\begin{definition}
 A subset of codewords with one finite non-zero trellis branch will be denoted as $\mathcal{C}_f$.  $\mathcal{C}_f$ consists of the codewords $\textbf{v}(D)$, that have finite non-zero trellis path of some length $j$, this path starts after initial zero state: $s_0=0$, $s_1 \neq 0$, does not have two consecutive zero trellis states in the non-zero branch, in zero state $s_j=0$ it merges with allzero path and the codeword does not have another non-zero trellis branch: $s_{j+1}=s_{j+2}=...=0$.
 \end{definition}

As codewords from the subset $\mathcal{C}_f$ have finite weight and this weight equals to the weight of a non-zero branch. We consider only this first non-zero part of codeword while defining active distances. The subset $\mathcal{C}_f$ does not include a zero codeword. The minimum length of non-zero trellis path for the codeword $\textbf{v}(D) \in \mathcal{C}_f$ is $m+1$, where $m$ is a memory of convolutional code. Here and after the trellis length is considered in terms of a number of output tuples or a number of transitions between trellis states. 

Let us denote the codeword from the subset $\mathcal{C}_f$ with non-zero trellis path of length $j$ by $\vec{v}^{(j)}$.

\begin{definition}
Active distance of length $j$ for the convolutional code is a minimum Hamming weight of the codewords $\textbf{v}^{(j)}$ from the set $\mathcal{C}_f$ that have non-zero trellis path of length $j$:
\begin{eqnarray*}
    a_j = \min_{\textbf{v}^{(j)}\in \mathcal{C}_f} w(\vec{v}^{(j)}) \ ,
\end{eqnarray*}
where the codeword $\vec{v}^{(j)} \in \mathcal{C}_f$, $w(\vec{v}^{(j)})$ is a Hamming weight of $\vec{v}^{(j)}$.
\end{definition}

Free distance of convolutional code can be also given in terms of active distance as considering subset $\mathcal{C}_f$ where we do not exclude codewords with smallest non-zero weight. Then the free distance is the smallest active distance over all $j>0$. It is important to note that the free distance may correspond to not the smallest length $j$. 

As the weight of the codeword $\vec{v}^{(j)} \in \mathcal{C}_f$ is affected only with first $j$ tuples, the calculation of a Hamming weight can be given as:
\begin{eqnarray*}
    w(\vec{v}^{(j)}) = \sum_{i=0}^{j-1} w(\vec{v}_{i}^{(j)}) \ .
\end{eqnarray*}

We denote by $N_{w^{(j)}}$ a number of codewords $\vec{v}^{(j)} \in \mathcal{C}_f$ with fixed weight $w^{(j)}$. Then we also present the definition of distance spectrum of the active distances that at first was suggested in \cite{isita_burst}.

\begin{definition}
Distance spectrum $\mathcal{D}_{a_j}$ of an active distances $a_j$ of length $j$ for convolutional code is a set of all possible weights of the codewords $\textbf{v}^{(j)}$ from the set $\mathcal{C}_f$ with non-zero trellis path of length $j$ and the numbers of codewords with that weights:
\begin{eqnarray*}
    \mathcal{D}_{a_j} = \{ (w^{(j)}, N_{w^{(j)}}) | w^{(j)} = w(\textbf{v}^{(j)}), \textbf{v}^{(j)} \in \mathcal{C}_f\} \ .
\end{eqnarray*}
\end{definition}

The minimum weight in the distance spectrum of the active distances of length $j$ is the active distance $a_j$ of length $j$ of the convolutional code. 

The minimum length $j$ which is feasible for the distance spectrum of the active distances for convolutional code with memory $m$ is $m+1$, as we have mentioned earlier. 

The union of all length forms a distance spectrum of active distances for the convolutional code.

\begin{definition}
Distance spectrum $\mathcal{D}_{a}$ of active distances for the convolutional code is a union of $\mathcal{D}_{a_j}$ for all possible non-zero trellis path lengths $j$:
\begin{eqnarray*}
    \mathcal{D}_a = \cup_{j=m+1}^{\infty} \mathcal{D}_{a_j} \ ,
\end{eqnarray*}
where $m$ is a code memory.
\end{definition}

\section{Bounds on Bit Error Rate}

Here we provide estimates for BER for the convolutional code with Viterbi decoding in the BSC. For this purpose let us first derive estimates on error burst probabilities.

\subsection{Probability of error burst}

The probability of the occurrence of an error burst and its estimations were discussed in \cite{isita_burst} and here we used the suggested there formulas.

First, we should calculate a probability of a fixed number of errors in a binary sequence. This probability is given as:
\begin{eqnarray*}
    P( e_{all}, e_1, s, w, p) =  \binom{w}{e_1} \binom{2s-w}{e_{all}-e_1}  p^{e_{all}} (1-p)^{2s-e_{all}} \ ,
\end{eqnarray*}
where $e_{all}$ is a number of total corrupted bits (number of bit errors), $e_1$ is a number of corrupted bits equal to one, $s$ is a length of the binary sequence in terms of tuples, $w$ is a hamming weight of the sequence, $p$ is a crossover probability of BSC.

Probability of an error burst of a given length $s$ and weight $w$ for binary symmetric channel with crossover probability $p$ is:
\begin{eqnarray*}
    P_{burst}(s, w, p) = \sum_{i>\frac{w}{2}}^{2s} \sum_{i_1 > \frac{w}{2}} ^ {min(w, i)}  P( e_{all}=i, e_1 = i_1, s, w, p) +
\end{eqnarray*}
\begin{equation}
    +\begin{cases}
    0 & \text{odd $w$,} \\
    \frac{1}{2}\sum_{i = \frac{w}{2}}^{2s-\frac{w}{2}}  P(e_{all}=i, e_1 = \frac{w}{2}, s, w, p)  & \text{even $w$.}  \label{main}
    \end{cases}
\end{equation}

Then lower and upper bounds on the probability of an error burst of length $j$ can be given with active distances $a_j$ and weight spectrum of active distances $\mathcal{D}_{a_j}$ for the code as in \cite{isita_fer}:

\begin{equation}
    P_{low}(s=j, p) = (1-p)^{2*2m}P_{burst}(s=j, w=a_j, p) \ , \label{plow}
\end{equation}

and 

\begin{equation}
    P_{up}(s=j, p) = \sum_{w_i=w_{min}}^{w_{max}} N_{w_i} P_{burst}(s=j, w = w_i, p) \ , \label{pup}
\end{equation}
where $ w_i, N_{w_i} \in \mathcal{D}_{a_j} $ and $m$ is a code memory.





\subsection{Bit Error Rate}

Bit error rate of convolutional code is a portion of the erroneous bits in the decoded codeword. Erroneous bits occurs only in the position of the error burst. The weight of an error burst corresponds to some weight from a distance spectrum of active distances. The minimal weight of an error burst for given length is an active distance for this length.

The estimate of the bit error rate can be given as an average portion of erroneous bits multiplied by the probability of error bursts and their weights. For lower and upper estimates it can be used lower and upper estimates for error burst probability correspondingly with multiplier $\frac{1}{n}$, where $n$ is a length of the tuple. In our case $n=2$. It is correct due to the fact that the probability of an error burst is the probability of its occurrence in a tuple, not in a single bit. Then for average portion of erroneous bits it has to be divided by the tuple length.

The most likely error bursts between all possible bursts have the minimum weight (since $p<1/2$) and correspond to the minimal active distance. As they have the greatest probability, these bursts make the greatest contribution to the BER. There can be more than one error burst with minimal weight, and they may have different lengths. These error bursts with minimum weight $w_{min} = \min_{j} a_j$ are equiprobable and the lower estimate of their probability can be given by $P_{low}(s=j_{w_{min}}, p)$, where $j_{w_{min}}$ can be any length $j$ that has corresponding minimal active distance $w_{min}$. The lower estimate of BER is suggested as the average portion of corrupted bits with only the most probable bursts:

\begin{equation}
    BER_{low}(p) = \frac{w_{min}N_{w_{min}}}{2} * P_{low}(s=j_{w_{min}}, p) \ , \label{plow}
\end{equation}
here $N_{w_{min}}$ is a number of the different most probable bursts, $ w_{min}, N_{w_{min}} \in \mathcal{D}_{a} $.

The upper estimate should take into account not only the most likely bursts but also bursts with greater weights. For the error burst probability it is used its upper estimate $P_{up}(s, p)$. For upper estimate we consider only the greatest possible weight from the spectrum of active distances $\mathcal{D}_{a_j}$ of the burst of length $j$. Then the upper estimate is a sum over all possible burst lengths' $j$:

\begin{equation}
    BER_{up}(p) = \min\left\{\sum_{j} \frac{\max_{w_j \in \mathcal{D}_{a_j}} w_j}{2} * P_{up}(s=j, p),1\right\} \ . \label{pup}
\end{equation}

\begin{figure}[htbp]
\centerline{\includegraphics[scale=0.3]{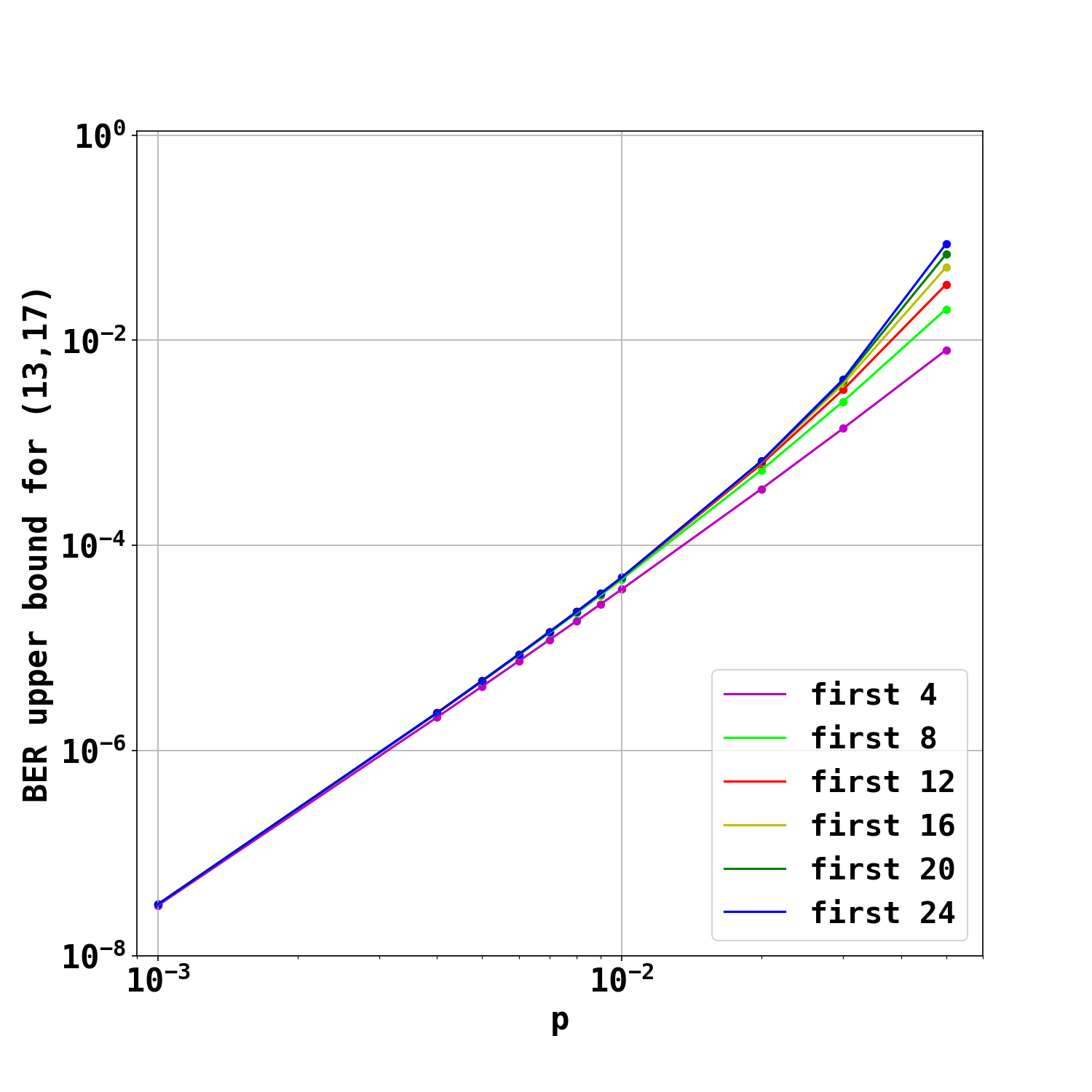}}
\caption{Dependence between $j$ and BER upper estimates for $(13,17)$ convolutional code with memory $m=3$}
\label{varL}
\end{figure}

In fact the probability of an error burst decreases exponentially with the weight of burst. That is why for a practical calculation of BER only first burst lengths $j$ can be used. We demonstrate this fact in Fig.~\ref{varL} where BER upper estimates were calculated for different $j$. 

From Fig.~\ref{varL} it can be noticed that difference between BER upper estimates obtained for different values of $j$, i.~e. for different active distances spectra decreases while $j$ increases. Moreover, any significant differences between estimates are observed only for high crossover probabilities of BSC where Monte Carlo technique can be applied to estimate BER of convolutional code. For the region of low probabilities $p$ ($p\leq 0.01$) all estimates almost coincide. It means that in this case small values of $j$ can be taken into account to estimate $BER_{up}(p)$. It means that $BER_{up}(p)$ can be estimated with low complexity.


\subsection{Simulation results}

\begin{figure}[htbp]
\centerline{\includegraphics[scale=0.3]{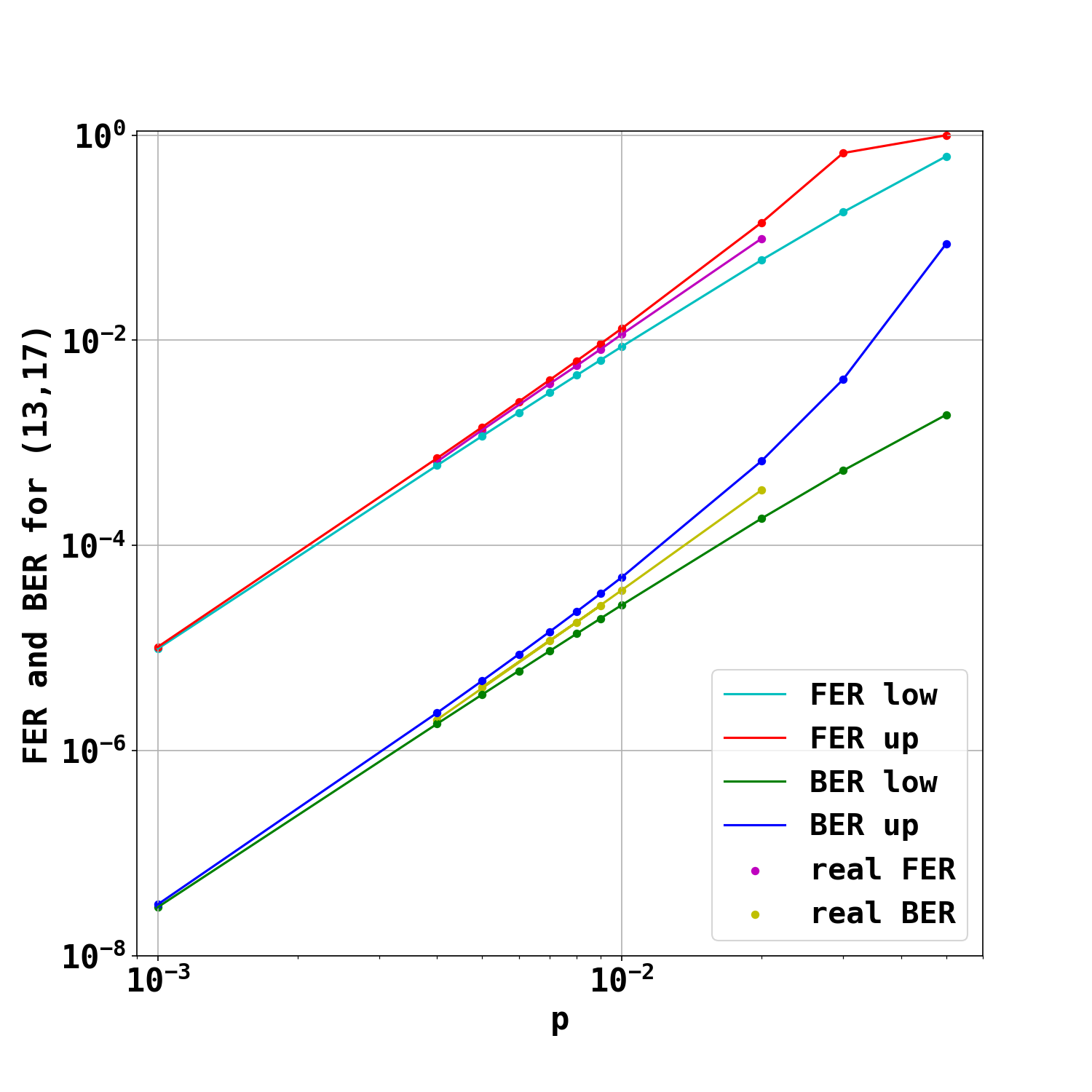}}
\caption{BER for convolutional code $(13,17)$ with memory $m=3$}
\label{ber1317}
\end{figure}

\begin{figure}[htbp]
\centerline{\includegraphics[scale=0.3]{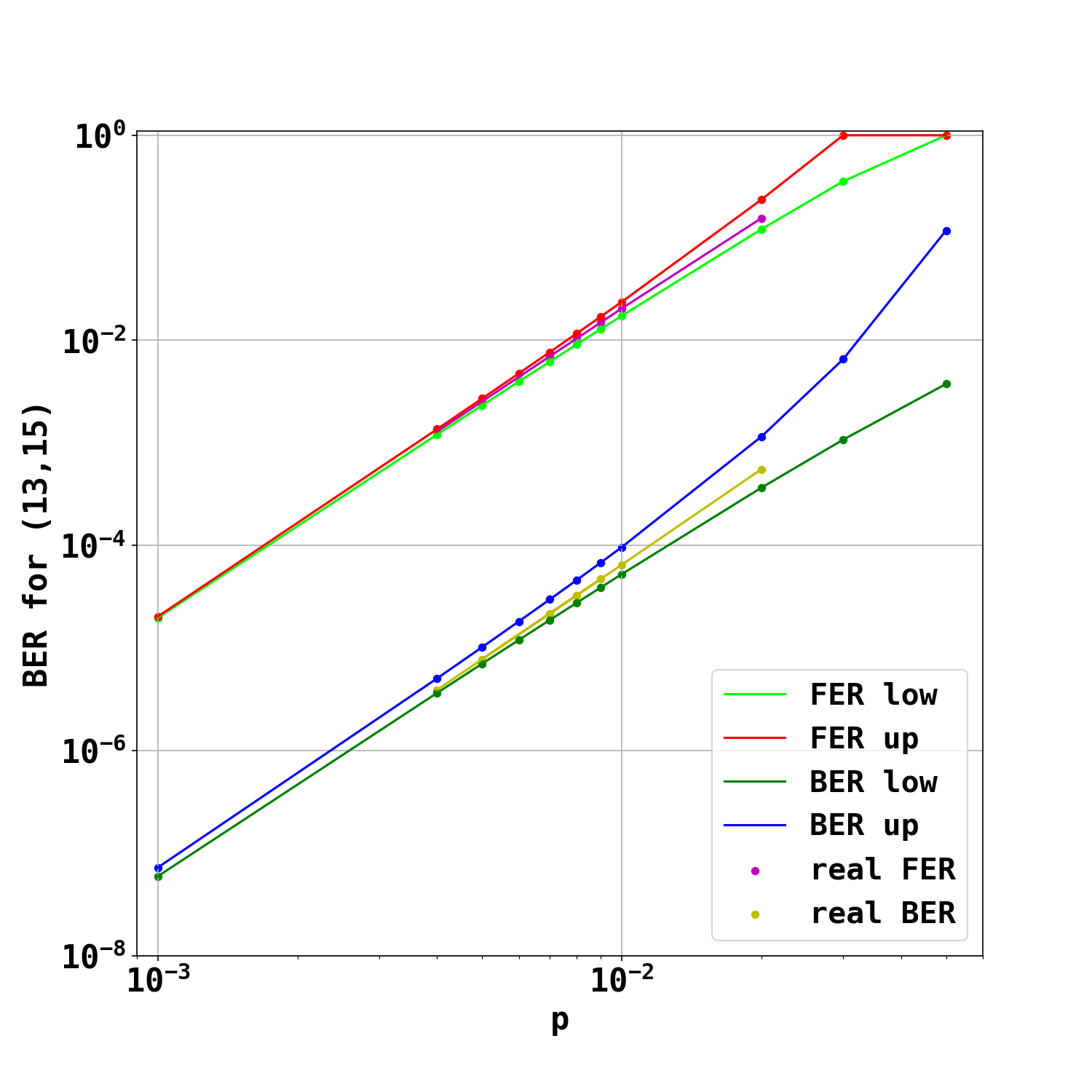}}
\caption{BER for convolutional code $(13,15)$ with memory $m=3$}
\label{ber1315}
\end{figure}

Here we present simulation results. In the simulations it was used binary symmetric channel with crossover probability $p$. Over the channel it was transmitted the finite random codewords that have length 1000 tuples (2000 bits). The convolutional codewords were zero terminated. In Fig. \ref{ber1317} and \ref{ber1315} BER and FER results for two codes $(13, 17)$ and $(13, 15)$ of memory $m=3$ are presented. 

The FER estimates are taken from \cite{isita_fer} and provided for a comparison. These results can be compared to the ones from \cite{bocharova2012closed} for code $(13,17)$ and it is clear that they are similar. Nevertheless, in \cite{bocharova2012closed} it was mentioned that it is impossible to find the BER analytically using their method for codes with memory greater than $4$. 

\begin{figure}[htbp]
\centerline{\includegraphics[scale=0.3]{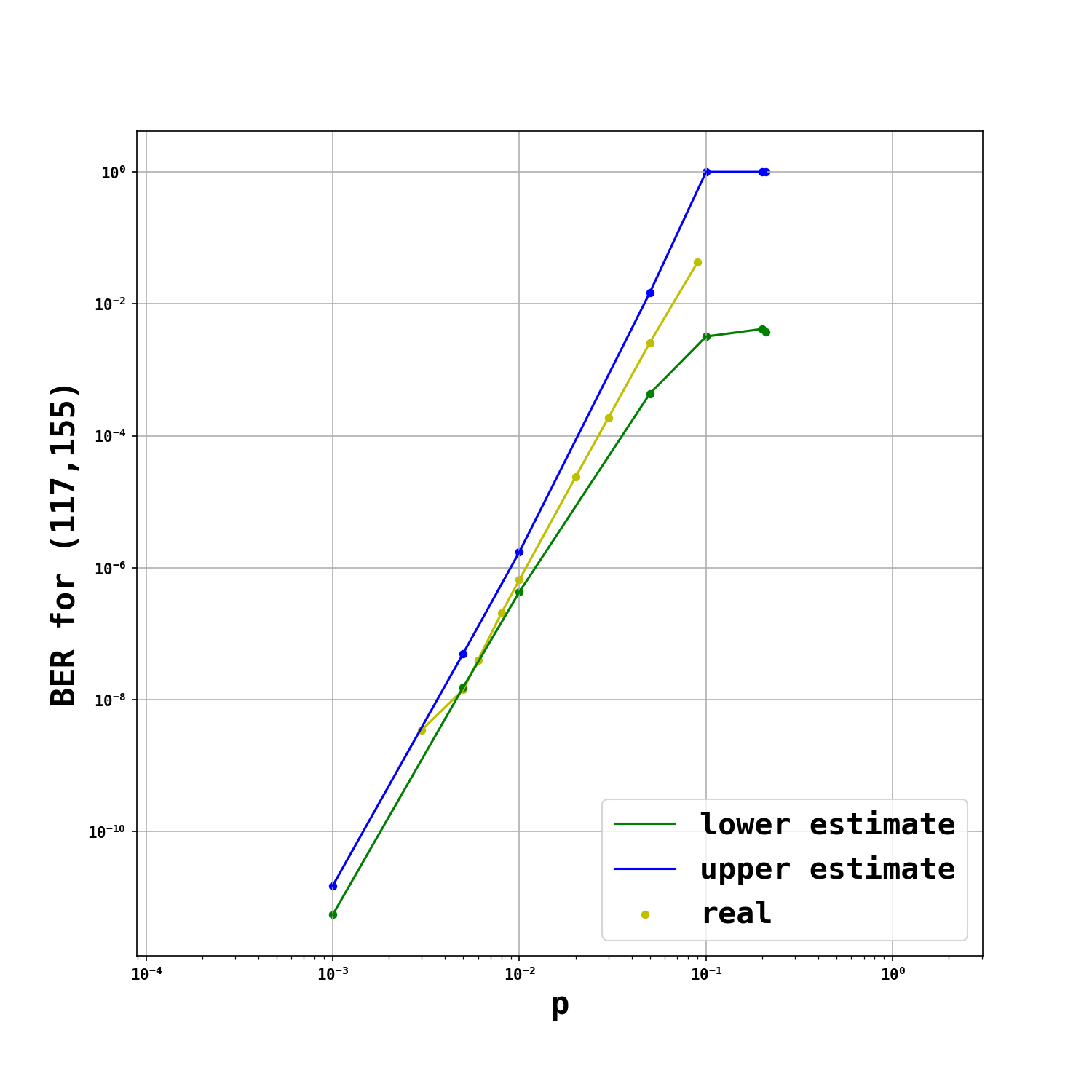}}
\caption{BER for convolutional code $(117,155)$ with memory $m=6$}
\label{ber117_155}
\end{figure}

The approach suggested in this work is based on the weight spectrum of active distances and the most computationally expensive procedure there is the active distance spectrum calculation. As for practical purposes it is enough to know the spectrum for the first lengths with lowest weights, and this computation is fast enough. The code for spectrum calculation and the results for described here codes can be found on Github \cite{git_see}. The estimates can be obtained easily with known code spectrum for any code memory $m$ and BSC crossover probability $p$. Here we also provide the BER performance and its lower and upper estimates for convolutional code $(117,155)$ of memory $m=6$ in Fig. \ref{ber117_155}. 

From provided results it is clear that the upper and lower estimates are close one to another for low crossover probabilities $p$. Thereby for these probabilities the estimates are precise while Monte-Carlo method is very time-consuming. It is also important to note that the spectrum of active distances is more efficient and faster to compute than the usual code spectrum as it is not taken into account the codewords that follow zero path.

\section{Conclusion}

In this paper it was suggested a method for constructing the upper and lower BER estimates for convolutional codes in BSC. These estimates are derived for maximum likelihood decoding of considered here systematic recursive convolutional code of rate $R=\frac{1}{2}$. The approach of BER estimation described in this paper uses the weight spectrum of active distances and has low complexity for any code memories and crossover probabilities. The results are provided here for different codes with different code memories and show that estimates are precise for crossover probabilities with practical use.

\section{Acknowledgement}

This work is an output of a research project implemented as part of the Basic Research Program at the National Research University Higher School of Economics (HSE University).

\bibliographystyle{IEEEtran}
\bibliography{IEEEexample}


\end{document}